\documentclass[a4paper,12pt]{article}

\usepackage{epsfig}
\usepackage{amssymb}
\usepackage{accents}

\newcommand{\ie}{{\it i.e.}}
\newcommand{\eg}{{\it e.g.}}
\newcommand{\mrm}[1]{\mbox{\rm #1}}
\newcommand{\be}{\begin{equation}}
\newcommand{\ee}{\end{equation}}
\newcommand{\br}{\begin{eqnarray}}
\newcommand{\bea}{\begin{eqnarray}}
\newcommand{\eea}{\end{eqnarray}}
\newcommand{\er}{\end{eqnarray}}
\newcommand{\ba}{\begin{array}}
\newcommand{\ea}{\end{array}}
\newcommand{\bi}{\begin{itemize}}
\newcommand{\ei}{\end{itemize}}
\newcommand{\bn}{\begin{enumerate}}
\newcommand{\en}{\end{enumerate}}
\newcommand{\bc}{\begin{center}}
\newcommand{\ec}{\end{center}}

\newcommand{\nn}{\nonumber\\}

\newcommand{\eq}[1]{eq.~(\ref{#1})}

\newcommand{\rfn}[1]{(\ref{#1})}

\newcommand{\effn}[1]{\accentset{(#1)}}

\newcommand{\gsim}{\lower.7ex\hbox{$\;\stackrel{\textstyle>}{\sim}\;$}}
\newcommand{\lsim}{\lower.7ex\hbox{$\;\stackrel{\textstyle<}{\sim}\;$}}

\begin{document}

\tolerance=100000
\thispagestyle{empty}
\setcounter{page}{0}

\begin{flushright}
CERN-PH-TH-2005/089\\
{\tt hep-ph/0506122}

\end{flushright}

\vspace*{\fill}

\begin{center}

{\Large \bf 

Running of Low-Energy Neutrino Masses, Mixing Angles and CP Violation}
\\[3.cm]

{  {\large\bf John Ellis$^1$, Andi Hektor$^2$,
Mario Kadastik$^2$, Kristjan Kannike$^3$,  Martti Raidal$^2$}}  
\\[7mm]

{\it
$^1$TH Division, PH Department, CERN, 1211 Geneva 23, Switzerland \\
$^2$National Institute of Chemical Physics and Biophysics, 
Ravala 10, Tallinn 10143, \\ Estonia \\
$^3$ Department of Physics and Chemistry, University of Tartu, Tahe 4, Tartu 51010, \\Estonia} 
\\[10mm]

\end{center}
 
\vspace*{\fill}

\begin{abstract}

{\small\noindent

We calculate the running of low-energy neutrino parameters from the bottom
up, parameterizing the unknown seesaw parameters in terms of the dominance
matrix $R$. We find significant running only if the $R$ matrix is
non-trivial and the light-neutrino masses are moderately degenerate. If
the light-neutrino masses are very hierarchical, the quark-lepton
complementarity relation $\theta_c + \theta_{12} = \pi/4$ is quite stable,
but $\theta_{13,23}$ may run beyond their likely future experimental
errors. The running of the oscillation phase $\delta$ is enhanced by the
smallness of $\theta_{13}$, and jumps in the mixing angles occur in cases
where the light-neutrino mass eigenstates cross.

}

\end{abstract}

\vspace*{\fill}

\begin{flushleft}

{\rm  June 2005} \\

\end{flushleft}

\newpage

\setcounter{page}{1}

\section{Introduction}

Low-energy neutrino experiments~\cite{nuexp} are providing crucial insight
into lepton masses and mixing, though this is still limited in its scope.
The most economical model for light neutrino masses is the seesaw
model~\cite{seesaw}, but even the minimal model with three heavy singlet
neutrinos contains a total of 18 parameters in the neutrino
sector~\cite{valle,EHLR}.  So far, neutrino experiments provide us with
measurements of only four of these~\cite{SV}: two squared-mass differences
$\Delta m^2_{12},$ $\Delta m^2_{23},$ and two neutrino mixing angles
$\theta_{23,12}$. There are prospects for measuring one more mixing angle,
$\theta_{13}$ and the CP-violating Maki-Nakagawa-Sakata phase~\cite{MNS}
$\delta$, as well, perhaps, as the overall neutrino mass scale in
cosmological data~\cite{astronu} and one combination of Majorana 
mass parameters in neutrinoless double-$\beta$ decay~\cite{beta}. 
However, even these measurements would
fall short of providing complete information on the full set of nine
parameters that are in principle observable in low-energy neutrino
experiments~\cite{DI}, out of the full total of 18.

Nevertheless, the information available from low-energy neutrino
experiments is already striking~\cite{SV}. The atmospheric mixing angle
$\theta_{23}$ is close to maximal: $\sin^2 2\theta_{23}=1.02\pm 0.04$, and
the solar mixing angle $\theta_{12}$ is quite large:  
$\tan^2\theta_{12}=0.45\pm 0.05$. It therefore seems that neutrino mixing
must be qualitatively different from the smaller mixing visible between
the left-handed quarks, where the largest mixing angle is the original
Cabibbo angle: $\sin\theta_{C}=0.22$. Such a difference in the quark and
neutrino mixing patterns was not widely expected before the experiments,
and has given rise to much theoretical discussion and
speculation~\cite{AF,GN,S}.

One of the problematic issues in the interpretation of the low-energy
neutrino data is the running of neutrino masses and mixing parameters
below and above the seesaw mass scales. This renormalization inevitably
introduces some `fuzziness' in the comparison between low-energy
measurements and any specific Ansatz for the mass matrix at the seesaw
scale, since the renormalization depends on many of the unknown parameters
in the seesaw model. The renormalization group equations (RGEs) have been
used to study this running extensively, both numerically and
analytically~\footnote{An up-to-date list of references on this very
extensive subject can be found in~\cite{A1}.}.  As a result, the observed
low-energy neutrino mixing can be obtained starting from either a
bimaximal~\cite{A2} or from an almost diagonal~\cite{HKL} neutrino mass
matrix at the Grand Unification (GUT) scale $M_{GUT}$.  
In such a situation, understanding the
systematical features of the running of neutrino parameters becomes
crucial for the interpretation of the neutrino data and for the building
of flavour models.

The purpose of this paper is to study comprehensively the dependence of
neutrino renormalization effects on {\it all} the seesaw parameters,
paying special attention to obtaining the correct low-energy neutrino
measurements. The running of the effective neutrino mass matrix below the
lightest singlet neutrino mass is generally well under control and large
renormalization effects can be expected only in the case of degenerate
light neutrino masses and, in supersymmetric models, for very large values
of $\tan\beta$~\cite{lola,A3}. However, understanding the renormalization
effects above and between the heavy neutrino scales~\cite{K,A1} is much more
complicated, since new non-trivial dependences on the heavy neutrino
Yukawa couplings $Y_\nu^{ij}$ are introduced. Because the flavour
structure of the new contribution to the RGEs can be very different from
that due to the effective neutrino mass matrix, large effects are
possible.  Since the couplings $Y_\nu^{ij}$ are largely unknown, a typical
{\it top-down} approach taken in previous studies has been to fix the
neutrino parameters at $M_{GUT}$ at some chosen values, then to run them
down to the electroweak scale and demonstrate that, for this particular
choice, the low-energy neutrino mass matrix is compatible with
experimental data.

In this paper we take a {\it bottom-up} approach in which we first fix the
known low-energy neutrino parameters to their measured values, and
evaluate renormalization towards higher scales consistently in the
framework of the minimal supersymmetric seesaw model. In our approach,
every set of studied neutrino parameters is physical by construction.  We
parameterize the nine high-energy parameters of the seesaw mechanism using
the orthogonal complex matrix $R$~\cite{CI}, and scan over all the 18
seesaw parameters by generating the unknown parameters (including phases)
randomly. We run the neutrino parameters up to the GUT scale and study the
dependence of the renormalization effects on $(i)$ the other observable
low-energy and $(ii)$ the high-energy parameters.

We find that significant renormalization effects can occur only when some
of the light neutrino masses get {\it comparable} contributions from two
or three heavy neutrinos $N_j$ and/or the light neutrino mass scale is at
least moderately degenerate.  Because the matrix $R_{ij}$, known as the
dominance matrix~\cite{masina}, characterizes the contributions of the
heavy neutrino $N_j$ to the light neutrino masses $\nu_i$, this
parametrization turns out to be quite useful for the present study. 
It has been stated in the literature that the solar angle
$\theta_{12}$ usually runs more than $\theta_{13,23}.$ We find that, for
light neutrino masses with a strong normal hierarchy, exactly the opposite
occurs. The quark-lepton complementarity relation~\cite{S,rode,R,comp2}
\bea
\label{prediction}
\theta_{C}+\theta_{12} & = & \frac{\pi}{4},
\eea
turns out to be very stable while, at the same time, the neutrino angles
$\theta_{13}$ and $\theta_{23}$ may receive renormalization effects larger
than the accuracy of plausible future experimental tests. The
renormalization of the low-energy oscillation phase $\delta$ 
is generally enhanced
compared with the running of mixing angles. Nevertheless, 
a non-diagonal  $R$-matrix is needed for a large effect also in this case.  
An interesting feature is the possible crossing of
light-neutrino mass eigenstates, which is accompanied by discrete changes
in the neutrino mixing pattern, and is correlated with the $R$-matrix
parameters.

Our paper is organized as follows. In Section 2 we present calculational
details of our study. In Section 3 we present and discuss our results.  
Finally, some conclusions are drawn in Section 4.

\section{Running of Neutrino Parameters in the MSSM}

The superpotential of the minimal supersymmetric standard model (MSSM)
with singlet (right-handed) heavy neutrinos is given by
\bea
W= D^c Y_d Q H_1 + U^c Y_u Q H_2 + E^c Y_e L H_1 + N^c Y_\nu L H_2 
+\frac{1}{2} N^c M N^c,  
\label{WMSSM}
\eea
where the Yukawa matrices $Y$ are general complex $3\times 3$ matrices and
the $3\times 3$ heavy neutrino mass matrix $M$ is symmetric.  The Yukawa
matrices can be diagonalized by bi-unitary transformations $Y^D=U^\dagger
Y V$, where $V,\,U$ refer to the rotation of the left- and right-chiral
fields, respectively. In the case of the symmetric matrix $M$, $U=V^*$. To
explain the neutrino data naturally, the hierarchy in $M$ should
preferably be of the same order as the square of the hierarchy in
$Y_\nu$~\cite{O}. We therefore assume hierarchical heavy-neutrino masses:
$M_{1} \leq M_{2} \leq M_{3}.$

Integrating out all the heavy singlet neutrinos, one gets the usual
dimension-5 effective operator
\bea
{\cal L}=-\frac{1}{2} \kappa LL H_2 H_2,
\eea
which after electroweak symmetry breaking gives masses to the light
neutrinos:
\bea
m_\nu(\mu)=\kappa(\mu) v^2 \sin^2\beta,
\eea
where $\mu$ is the renormalization scale, $v=174$ GeV and
$\tan\beta=v_2/v_1$ is the ratio of the v.e.v.'s of the corresponding
Higgs doublets. Above the heaviest neutrino mass scale, $\mu>M_3,$ the
light-neutrino mass matrix reads
\bea
m_\nu(\mu)=Y_\nu^T(\mu) M^{-1}(\mu)Y_\nu(\mu) v^2 \sin^2\beta.
\label{mnu}
\eea
Between the heavy-neutrino mass scales, $M_{1} \leq M_{2} \leq M_{3},$
there exist a series of effective theories with, in general, $n$ active
heavy neutrinos. The tree-level matching conditions between these theories
at the neutrino thresholds are
\begin{equation}
\label{match}
\effn{n} \kappa_{gf}|_{M_{n}} = \effn{n+1} \kappa_{gf}|_{M_{n}} + 
(\effn{n+1}{Y_\nu})_{ng} \effn{n+1}M^{-1}_n 
(\effn{n+1}Y_\nu)_{ng}|_{M_{n}},
\end{equation}
where $(n)$ is the number of heavy neutrinos not integrated out. In
general, the light-neutrino mass matrix can be written as
\bea
m_\nu= \left( \effn{n} \kappa + 
\effn{n}{Y_\nu^T} \effn{n}M^{-1} \effn{n}Y_\nu \right) v^2 \sin^2\beta.
\label{mnu2}
\eea
Since $m_\nu$ and $Y_e^\dagger Y_e$ can be diagonalized with the unitary
matrices $V_\nu$ and $V_e$, respectively, the mixing matrix observable in
the low-energy experiments is
\bea
V_{MNS}=V^\dagger_e V_\nu .
\label{VMNS}
\eea                           

While $m_\nu$ contains 9 physical parameters, $Y_\nu$ and $M$ together
contain 18 parameters. The missing 9 parameters crucially affect the
physical observables. These include, for example, renormalization-induced
lepton-number-violating processes~\cite{BM,Hisano,CI,EHLR,ER} and
electric dipole moments~\cite{EHLR,EHRS} in the supersymmetric seesaw model, 
as well as the renormalization of the light-neutrino parameters~\cite{A1}.
Therefore, to study the dependence of the renormalization of \eq{mnu2} on
$Y_\nu$, we parametrize $Y_\nu$ with the complex orthogonal matrix
$R$~\cite{CI}. In the basis in which $M$ is diagonal, we write
\bea
\effn{n} Y_\nu= (\effn{n} M^D)^\frac{1}{2} \;\effn{n} R \; 
(m_\nu^D)^\frac{1}{2}\;V_\nu^\dagger\;
(v \sin\beta)^{-1},
\label{R}
\eea
where the matrix $R$ is parametrized in terms of three \emph{complex}
angles $\theta_{12}^R$, $\theta_{13}^R$ and $\theta_{23}^R$:
\begin{equation}
\label{eq:R:parametr}
R
= \left(
\begin{array}{ccc}
c_{12}^R c_{13}^R & s_{12}^R c_{13}^R & s_{13}^R  \\
-c_{23}^R s_{12}^R - s_{23}^R s_{13}^R c_{12}^R  & c_{23}^R c_{12}^R - 
s_{23}^R s_{13}^R s_{12}^R  & 
s_{23}^R c_{13}^R  \\
s_{23}^R s_{12}^R - c_{23}^R s_{13}^R c_{12}^R  & -s_{23}^R c_{12}^R - 
c_{23}^R s_{13}^R s_{12}^R  & c_{23}^R c_{13}^R
\end{array} \right),
\end{equation}
where $s_{ij}^R \equiv \sin{\theta_{ij}^R}$ and $c_{ij}^R \equiv
\cos{\theta_{ij}^R}$. Since $Y_\nu$ and $M$ are renormalized, obviously
also $\effn{n} R$, which consists of $n$ rows, runs with energy.  The RGEs
for $Y_\nu$ and $M$ can be found in~\cite{Hisano}, and the RGEs for $R$
were calculated in~\cite{CEPRT}. Using these, $\effn{n} R$ has to be
evaluated at every heavy neutrino threshold when the matching is
performed.

The scale dependence of the effective/combined quantities in \rfn{mnu2} is
characterized by the differential equation~\cite{Aren,A1}
\bea
\label{eq}
16 \pi^{2} \frac{d \effn{n} X}{dt} &=& 
( Y_e^\dagger Y_e  + \effn{n} Y_\nu^\dagger \effn{n} Y_\nu)^T \effn{n} 
X + 
\effn{n} X ( Y_e^\dagger Y_e  + \effn{n} Y_\nu^\dagger \effn{n} 
Y_\nu)^T + 
\nn
&& 
  (2 \mrm{Tr} (\effn{n} Y_\nu^\dagger \effn{n} Y_\nu + 
3 Y_u^\dagger Y_u) -6/5 g_1^2 -6 g_2^2) \effn{n} X, 
\eea
where $X=\kappa, Y_\nu^T M^{-1} Y_\nu.$ Notice that below the $M_1$ scale
$\effn{n} Y_\nu=0.$ Therefore one expects large renormalization effects to
occur above the heavy-neutrino thresholds for two reasons. First, the
Yukawa couplings $Y_\nu$ can be large. Secondly, the flavour structure of
$Y_\nu^\dagger Y_\nu$ can be very different from the flavour structure of
$Y_e^\dagger Y_e$ and $\kappa.$ Both those effects can be traced back to
the values of $R$ via \eq{R}. Approximate analytical solutions to \eq{eq}
have been derived in~\cite{A1}, which allow one to understand the generic
behaviour of the renormalization effects. However, due to
enhancement/suppression factors and possible cancellations, the exact
numerical solutions may differ considerably from those estimates.

\section{Results for Normally-Ordered Light Neutrinos}

In this Section we present the results of our study for the case of
normally-ordered light-neutrino masses, using the following strategy. We
start at $M_Z$, where we fix the measured light-neutrino parameters as
$\Delta m^2_{12}=8.\times 10^{-5} \mrm{eV}^2,$ $\Delta m^2_{23}=2.2 \times
10^{-3} \mrm{eV}^2,$ $\tan^2\theta_{12}=0.41,$ $\sin\theta_{23}=0.7$ and
$\sin\theta_{13}=0.05.$ We then generate randomly the lightest neutrino
mass, the heavy neutrino masses, all the low-energy phases and the initial
values for the parameters in the $R$ matrix. We run all the relevant
quantities up to $M_{GUT}$ using the 1-loop RGEs for the minimal
supersymmetric seesaw model~\cite{Hisano,Aren}. At every heavy-neutrino
threshold we perform the tree-level matching according to \eq{match}. To
calculate the values of $\effn{n} Y_\nu$ we use the renormalized values of
the $R$ and $M$. At $M_{GUT}$ we calculate the renormalized light-neutrino
parameters. We always keep the ordering of the light neutrino masses fixed,
$m_1<m_2<m_3$ for the normal and $m_3<m_1<m_2$ for the inverse ordering.
Because of that the physical range for $\theta_{12}$ extends up to $\pi/2.$ 

In order to accentuate the renormalization effects due to the low-energy
neutrino parameters and the parameter matrix $R,$ we do not consider
degenerate light neutrinos and we assume an upper limit $m_1<0.1$ eV on the
lightest neutrino mass. Although the present most stringent limit on the
overall light-neutrino mass scale scale coming from astrophysics and
cosmology~\cite{WMAP} is a factor of 2 to 3 weaker, such precision can
easily be achieved in the future cosmological experiments~\cite{astronu}.
We also minimize the renormalization effects of large $\tan\beta$ 
studied in~\cite{A3} by
working with the relatively small value of $\tan\beta=5.$ Instead, we
study how the large values of $Y_\nu$ affect the renormalization effects.

\subsection{Renormalization of the Mixing Angles}

\begin{figure}[htbf]
\centerline{\epsfxsize = 0.45\textwidth  \epsffile{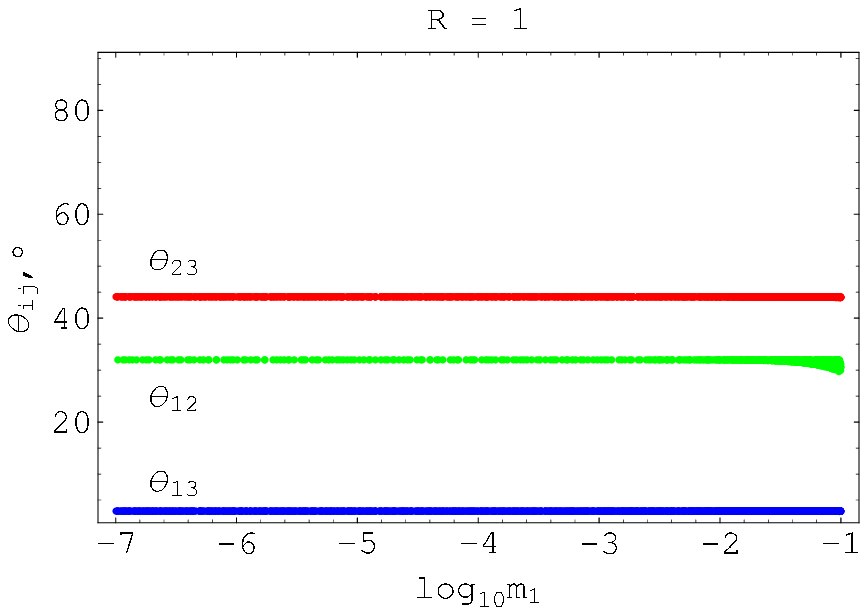}
\hfill\epsfxsize = 0.45\textwidth  \epsffile{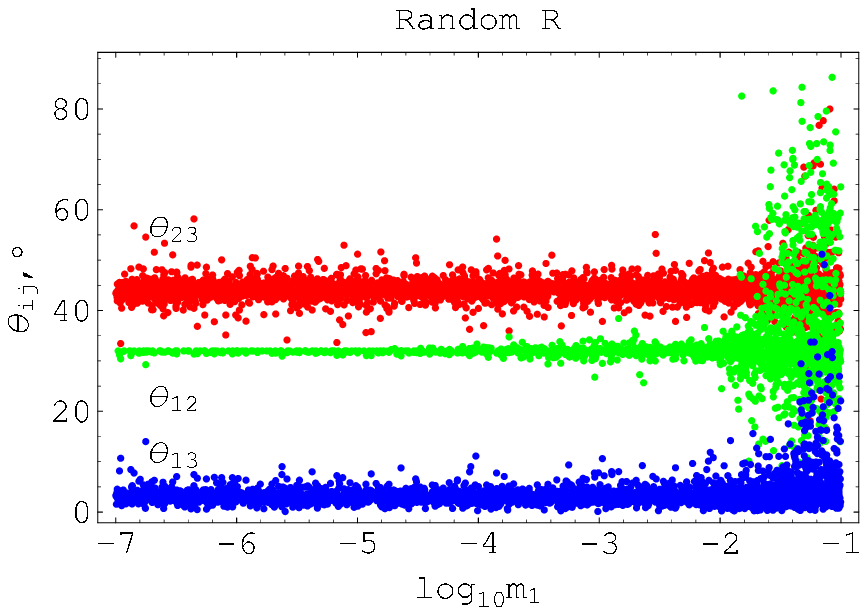}}
\caption{\it Neutrino mixing angles at $M_{GUT}$ as functions of the 
lightest
neutrino mass at $m_{1}(M_Z)$ for $R(M_Z)=1$ (left panel) and 
randomly generated $R$ (right panel). 
\vspace*{0.cm}}
\label{f1}
\end{figure}

We start by studying the running of the light-neutrino mixing angles. In
Fig.~\ref{f1} we plot the neutrino mixing angles $\theta_{ij}$ at
$M_{GUT}$ as functions of the lightest neutrino mass $m_{1}(M_Z)$ for
$R(M_Z)=1$ (left panel) and for randomly generated $R$ (right panel). 
In all the figures the neutrino mass parameters are presented in units 
of $eV.$ The mixing angles $\theta_{12},$ $\theta_{13}$ and $\theta_{23}$ 
are represented by green (light), blue (dark) and red (medium) dots,
respectively. For $R=1$ the mixing angles practically do not run: only
$\theta_{12}$ may change a little for light-neutrino masses close to
$m_1=0.1$ eV.  This is because, for moderate degeneracy, the
renormalization of $\theta_{12}$ is enhanced by a factor $m_1^2/\Delta
m_{12}^2.$ This effect would have been larger for larger values of
$\tan\beta$ and $m_1.$

Turning to the results for the randomly-generated values of $R$, a certain
pattern of renormalization effects emerges.  If $m_{1}>\sqrt{\Delta
m_{12}^2},$ very large changes in the mixing angles may occur. Although
$\theta_{12}$ tends to change most due to the above-mentioned enhancement
factor, also $\theta_{13}$ and $\theta_{23}$ may gain almost any value.  
We see that the examples of extreme running considered in the literature
are due to having at least a moderately degenerate mass spectrum.

An interesting set of points in Fig.~\ref{f1} are those gathered around
$\theta_{12}\sim 60^\circ$ in the region $m_{1}>\sqrt{\Delta m_{12}^2}.$
Those correspond to the level crossing of two light-neutrino mass
eigenvalues $m_1$ and $m_2$ due to renormalization. Because by definition 
$m_1<m_2,$ this causes a discrete
jump in the value of $\theta_{12}$. In the standard parametrization of
$V_{MNS}$, and with small $\theta_{13}$, this implies
$\sin\theta_{12}\to\cos\theta_{12}'$ and, consequently,
$\theta_{12}'=90^\circ -\theta_{12}.$ As seen in Fig.~\ref{f1}, this
effect is smeared by strong running of $\theta_{12}$ and also
$\theta_{13}.$

In contrast to the previous discussion, if the mass spectrum is strongly
hierarchical:  $m_{1}<\sqrt{\Delta m_{12}^2},$ the solar angle is much
more stable than the mixing angles $\theta_{13}$ and $\theta_{23}.$ The
latter may vary through a range of almost $10^\circ,$ which is more than
the expected precision of future experiments.  Moreover, the widths of the
$\theta_{13}$ and $\theta_{23}$ bands in Fig.~\ref{f1} do not depend on
the initial values of the angles $\theta_{ij}.$ Thus, {\it discrimination
between different flavour models may be possible in principle in the
future}, if one takes into account renormalization effects. We also note
that, for hierarchical light-neutrino masses, the quark-lepton
complementarity relation $\theta_{C}+\theta_{12} = \pi/4$ would be
maintained with high accuracy at every scale, independently of the
unobservable neutrino parameters.

\begin{figure}[t]
\centerline{\epsfxsize = 0.45\textwidth  \epsffile{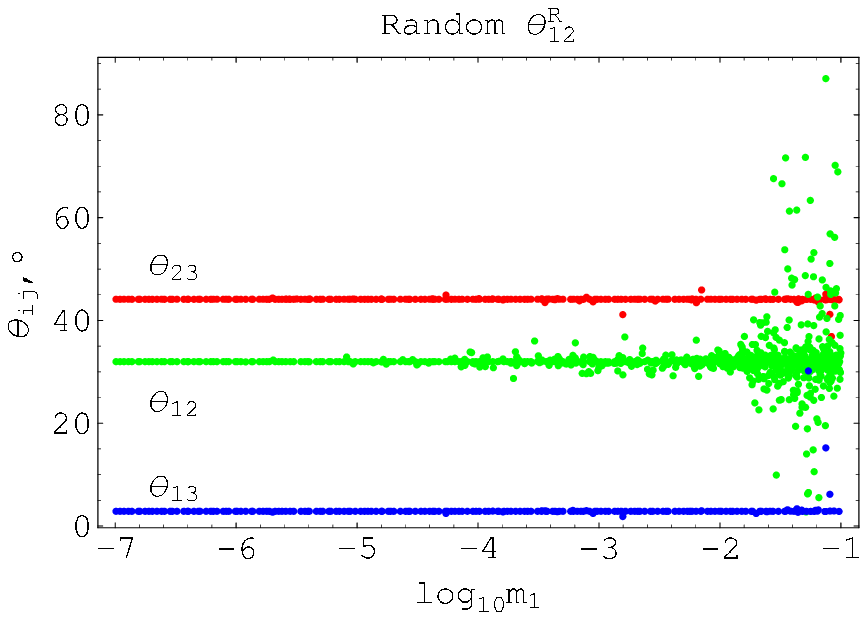}
\hfill \epsfxsize = 0.45\textwidth  \epsffile{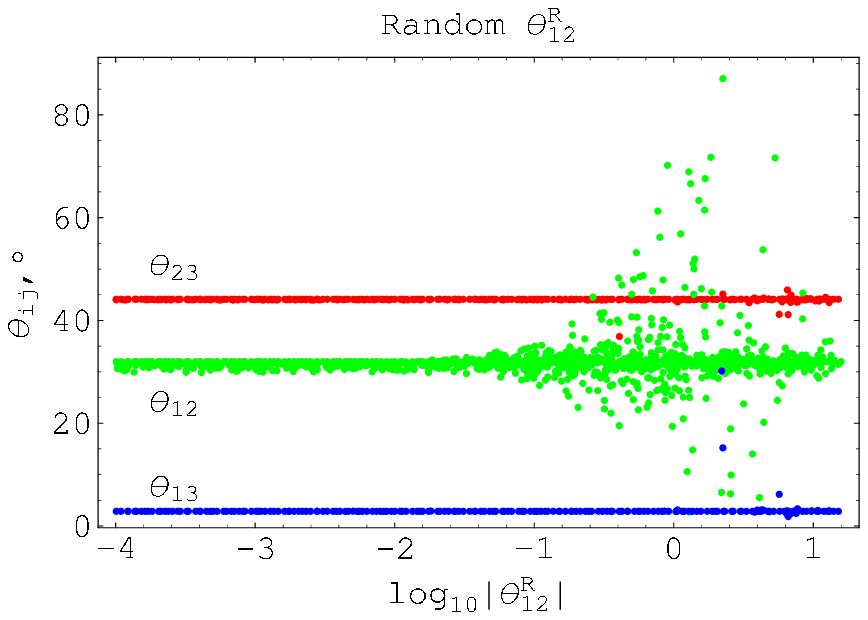}}
\caption{\it Neutrino mixing angles at $M_{GUT}$ as functions of the 
lightest
neutrino mass $m_{1}(M_Z)$ for random $\theta_{12}^R$ (left panel), 
and as functions of $|\theta_{12}^R|$ (right panel). 
The rest of the parameters in $R(M_Z)$ vanish. 
\vspace*{0cm}}
\label{f2}
\end{figure}

\begin{figure}[htbf]
\centerline{\epsfxsize = 0.45\textwidth  \epsffile{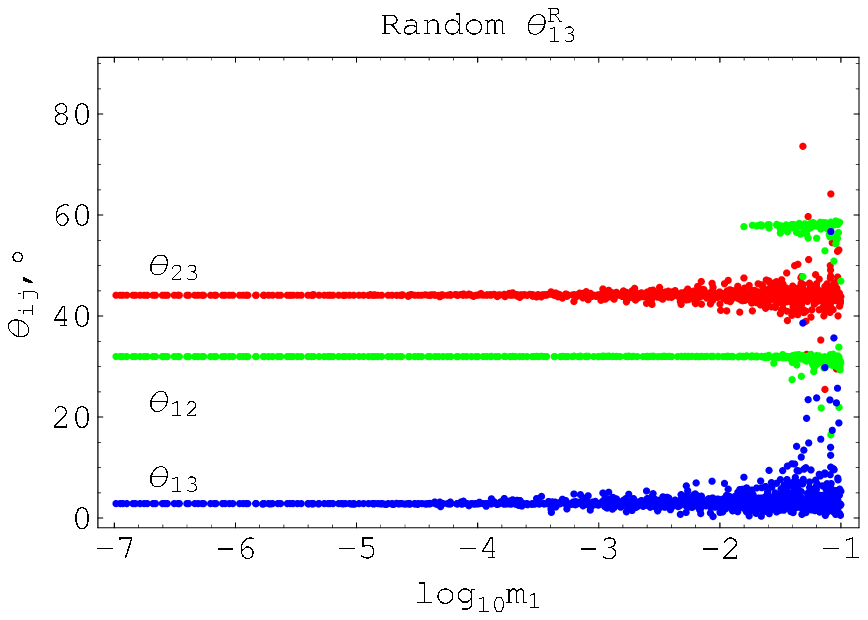}
\hfill \epsfxsize = 0.45\textwidth  \epsffile{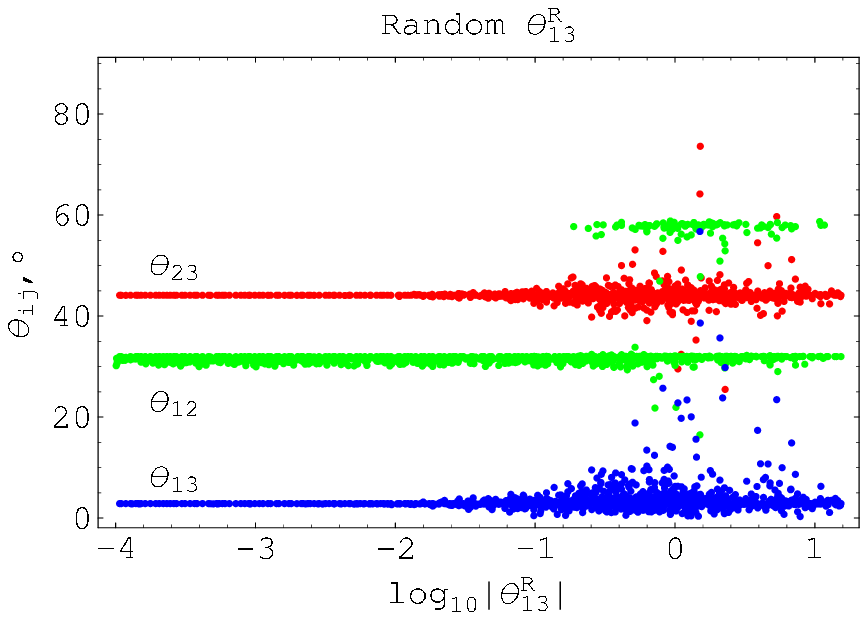}}
\caption{\it Neutrino mixing angles at $M_{GUT}$ as functions of the 
lightest
neutrino mass $m_{1}(M_Z)$ for random $\theta_{13}^R$ (left panel), 
and as functions of $|\theta_{13}^R|$ (right panel). 
The rest of the parameters in $R(M_Z)$ vanish. 
\vspace*{0cm}}
\label{f3}
\end{figure}

\begin{figure}[htbf]
\centerline{\epsfxsize = 0.45\textwidth  \epsffile{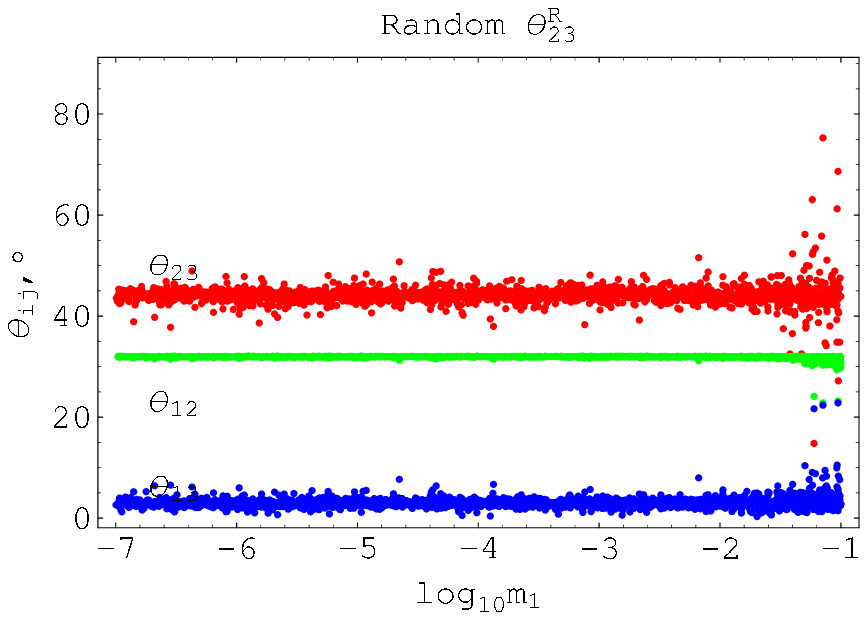}
\hfill \epsfxsize = 0.45\textwidth  \epsffile{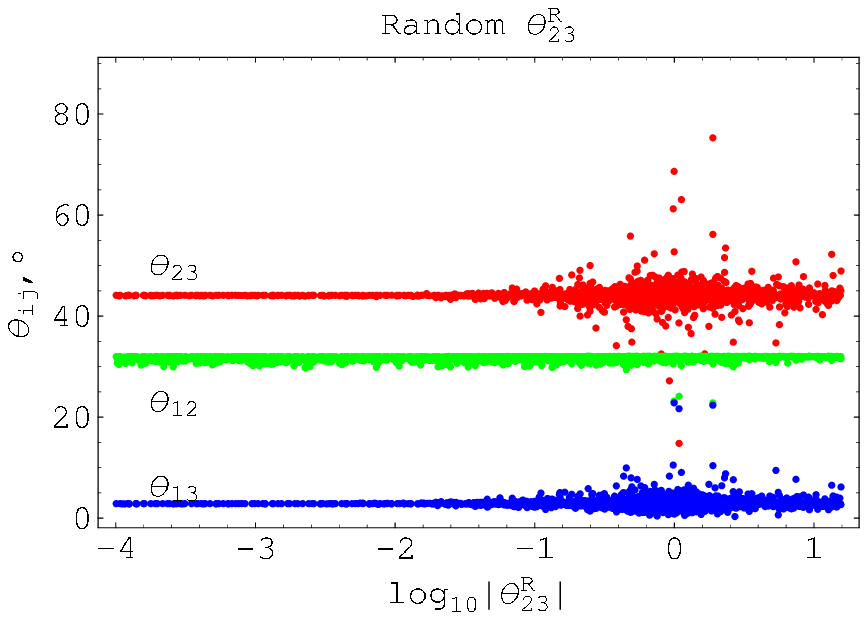}}
\caption{\it Neutrino mixing angles at $M_{GUT}$ as functions of the 
lightest
neutrino mass $m_{1}(M_Z)$ for random $\theta_{23}^R$ (left panel), 
and as functions of $|\theta_{23}^R|$ (right panel). 
The rest of the parameters in $R(M_Z)$ vanish. 
\vspace*{0cm}}
\label{f4}
\end{figure}

We now study the origins of the effects due to $R$. In the left panels of
Figs.~\ref{f2}, \ref{f3} and \ref{f4} we plot the distributions of the
neutrino mixing angles $\theta_{ij}(M_{GUT})$ as functions of $m_1(M_Z)$
for randomly generated complex parameters $\theta_{12}^R,$ $\theta_{13}^R$
and $\theta_{23}^R,$ respectively.  In each figure the other parameters in
$R$ are set to zero. The same neutrino mixing angles are plotted in the
right panels of Figs.~\ref{f2},~\ref{f3},~\ref{f4} as functions of the
absolute values of the corresponding $R$ matrix parameters
$|\theta_{12}^R|,$ $|\theta_{13}^R|$ and $|\theta_{23}^R|,$ respectively.

We see in Fig.~\ref{f2} that non-zero values of $\theta_{12}^R$ affect
mostly the renormalization of $\theta_{12}.$ A large running effect
requires also that the overall light neutrino mass scale be high.  On the
other hand, non-zero values of $\theta_{13}^R$ affect mostly the running
of $\theta_{13}$ and $\theta_{23},$ as seen in Fig.~\ref{f3}. Again, a
relatively high overall light-neutrino mass scale is required for a
significant effect. We also learn from Fig.~\ref{f3} that the level
crossing of light mass eigenvalues is induced by non-zero $\theta_{13}^R,$
which strongly affects the running of $m_1.$ The parameter $\theta_{23}^R$
affects only the running of $\theta_{13}$ and $\theta_{23}.$ Fig.~\ref{f4}
shows an interesting feature: in this case the running of $\theta_{13}$
and $\theta_{23}$ does not depend on $m_1,$ and significant
renormalization effects can be obtained also for very hierarchical light
neutrinos.

Comparison of the left and right panels in Figs.~\ref{f2}, \ref{f3} and
\ref{f4} reveals how the renormalization effects depend on the magnitude
of the particular parameter $\theta_{ij}^R.$ Interestingly, in all the
cases the dominant running occurs in the region $|\theta_{ij}^R|\sim {\cal
O}(1).$ We recall that $R$ can be interpreted to be a dominance
matrix~\cite{masina}, \ie, it shows which heavy neutrino contribution
dominates in the mass of a particular light neutrino. Therefore our
results imply that, in order to have significant renormalization effects,
{\it at least two heavy neutrinos must contribute to one particular light
neutrino mass in approximately equal amounts}. If the light neutrino
masses are dominated by one heavy neutrino each, no large running is
possible unless the light neutrinos are degenerate in mass and $\tan\beta$
is large.


\subsection{Renormalization of Masses}

The observed hierarchy in the light-neutrino masses, $\sqrt{\Delta
m^2_{12}/\Delta m^2_{23}}=0.18,$ is milder than expected in many flavour
models. In GUTs with the simplest scalar sector, the neutrino Yukawa
couplings are equal to the up-quark Yukawa couplings at $M_{GUT}.$
Contrary to that, phenomenology at the low scale seems to indicate that
the neutrino hierarchy is more similar to the less hierarchical
down-sector Yukawa couplings, rather than those in the up sector.  
However, at the moment the lightest neutrino mass and the hierarchy in the
heavy-singlet sector are unknown, introducing large uncertainties into
such considerations. Therefore it is interesting to study also how the
masses of the light neutrinos evolve with energy.

\begin{figure}[t]
\centerline{\epsfxsize = 0.45\textwidth  \epsffile{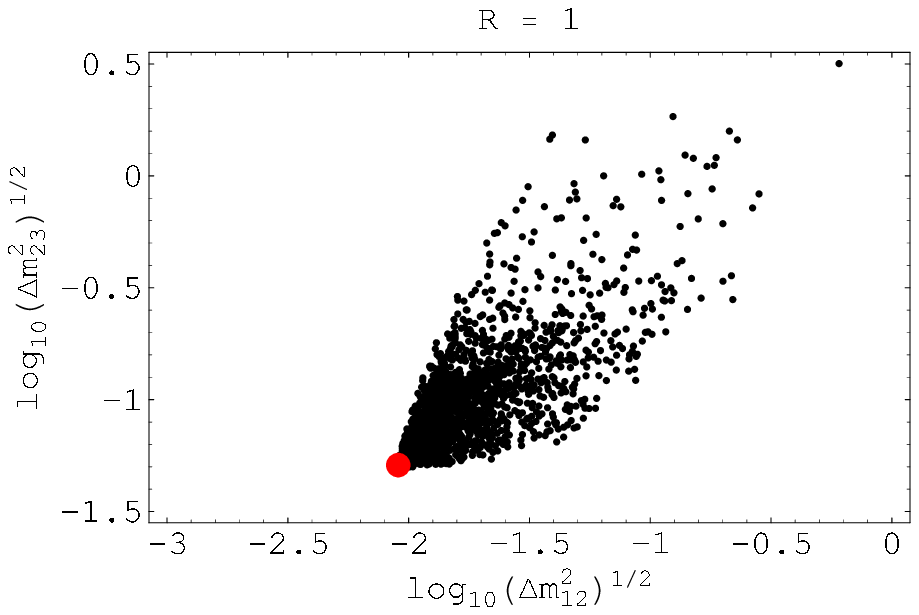}
\hfill\epsfxsize = 0.45\textwidth  \epsffile{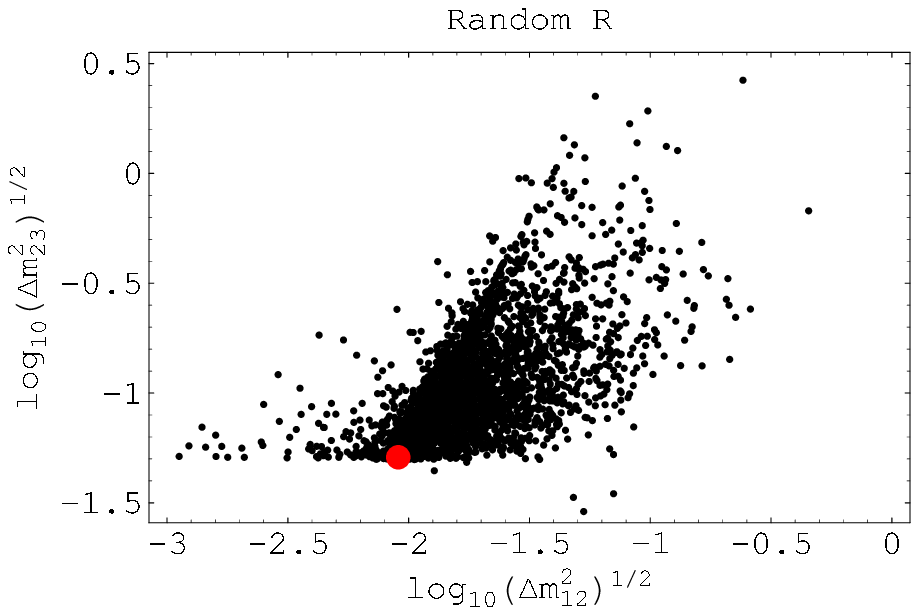}}
\caption{\it 
The distribution of $\sqrt{\Delta m_{23}^2}$ versus 
$\sqrt{\Delta m_{12}^2}$ at  $M_{GUT}$ for $R(M_Z)=1$ (left panel)
and for randomly generated $R$ (right panel).
\vspace*{0.cm}}
\label{f5}
\end{figure}

In Fig.~\ref{f5} we plot the distributions of $\sqrt{\Delta m^2_{12}}$ and
$\sqrt{\Delta m^2_{23}}$ at $M_{GUT}$ for $R(M_Z)=1$ (left panel) and for
randomly generated $R$ (right panel). The red dot denotes the starting
point at $M_Z$ from which value all the other points are generated.  {\it
The hierarchy at the GUT scale tends to be larger than at low energies},
although the opposite is also possible for a few points. While for $R=1$
both mass differences tend to increase, for random $R$ they may also
decrease. The abundant points with smaller values of $\sqrt{\Delta
m^2_{12}}$ in the right-hand plot correspond to non-zero values of
$\theta_{13}^R.$ This parameter affects the Yukawa couplings of first
generation in such a way that the $(12)$ mass difference may run
considerably.

\subsection{Renormalization of CP Observables}

Our approach in this study is to fix the known neutrino parameters and to
vary the unknown ones randomly. At the moment, the only CP-violating
observable in the neutrino sector what we have information about is the
baryon asymmetry of the Universe, assuming that the cosmological baryon
asymmetry is generated via leptogenesis~\cite{lepto}. In this case, it is
possible to constrain one combination of the 6 CP-violating phases in the
neutrino sector. However, because one can vary the remaining 5
combinations (and also the unknown CP-conserving neutrino parameters), one
cannot make any firm prediction for the neutrino parameters~\cite{ER}.
As the predictions for other renormalization-induced CP-violating
observables such as the electric dipole moments of the charged leptons are
orders of magnitude smaller than the present experimental
bound~\cite{EHLR,EHRS}, no firm constraints come from this sector either.
Although Ref.~\cite{A3} argues that some systematic renormalization
effects in leptogenesis
are possible due to the running of the effective light neutrino
mass matrix, those are already taken into account in the systematic study
of~\cite{GNRRS}.

\begin{figure}[t]
\centerline{\epsfxsize = 0.45\textwidth  \epsffile{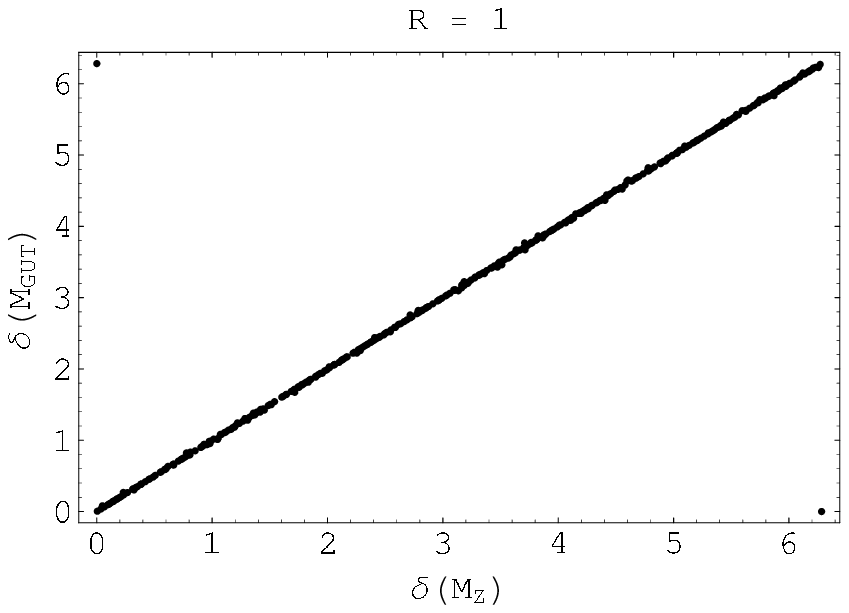}
\hfill\epsfxsize = 0.45\textwidth  \epsffile{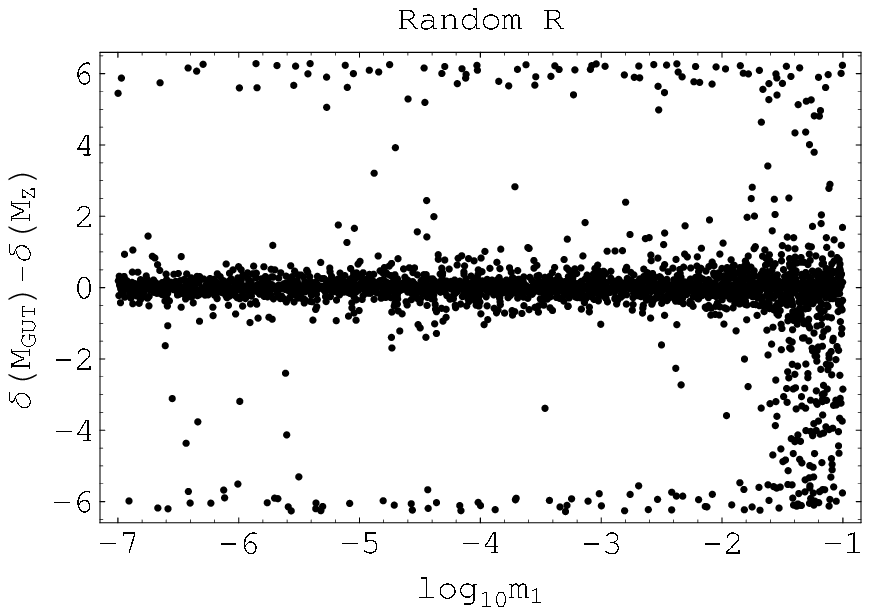}}
\caption{\it The MNS phase at high energy $\delta(M_{GUT})$ 
as a function of its low-energy value $\delta(M_Z)$ for $R(M_Z)=1$ (left 
panel), and $\delta(M_{GUT})-\delta(M_Z)$ 
as a function of $m_1$ for randomly generated $R$ (right panel). 
\vspace*{0.cm}}
\label{f6}
\end{figure}

At the moment, the most realistic possibility seems to be that of
measuring the MNS phase $\delta$ in future oscillation experiments. The
value of this phase, however, is presently unknown. In the left plot of
Fig.~\ref{f6} we present the values of $\delta(M_{GUT})$ against the
initial values of the phase for $R(M_Z)=1$. As seen in the figure there is
practically no running of $\delta$ in this case. The situation changes
considerably for randomly generated $R$-matrices, as seen in the right
panel of Fig.~\ref{f6} where we plot the change of the phase,
$\delta(M_{GUT})-\delta(M_Z),$ as a function of the lightest neutrino mass
$m_1.$ The running of the MNS phase can be numerically significant and,
apart from high values of $m_1,$ almost independent of the lightest
neutrino mass. This behaviour resembles the running of $\theta_{13,23}$ in
Fig.~\ref{f1} and can be traced back to the $1/\theta_{13}$ enhancement of
the running of $\delta$~\cite{A1}~\footnote{The points around $\pm 2\pi$
in Fig.~\ref{f6} actually correspond to small running modulo $2 \pi$.}.

\section{Renormalization Effects for an Inverted Hierarchy of Light
Neutrino Masses}

\begin{figure}[t]
\centerline{\epsfxsize = 0.45\textwidth  \epsffile{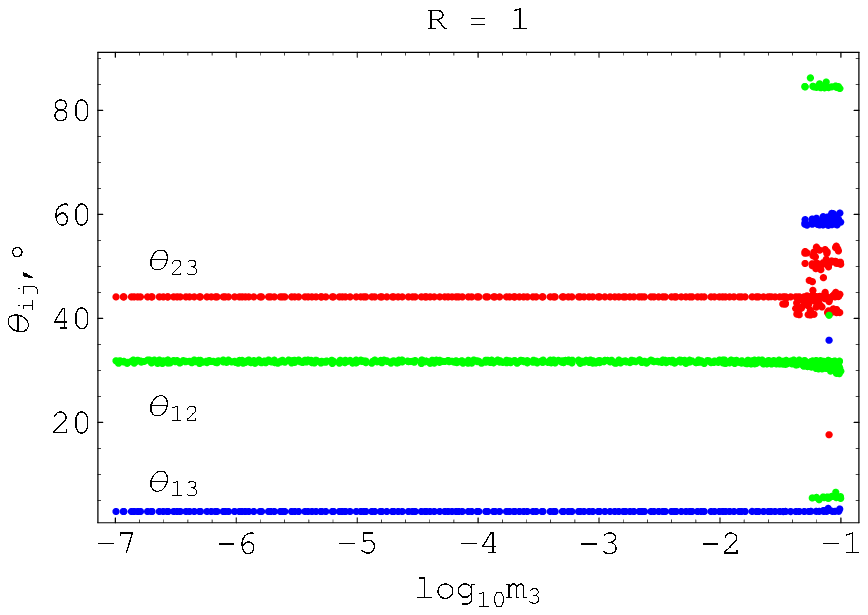}
\hfill\epsfxsize = 0.45\textwidth  \epsffile{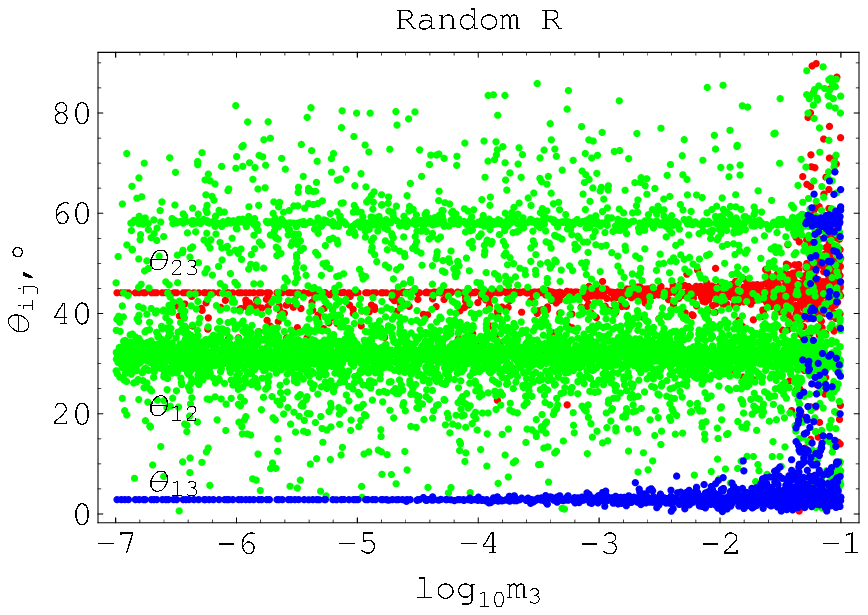}}
\caption{\it Neutrino mixing angles at $M_{GUT}$ as functions of the 
lightest
neutrino mass $m_{3}(M_Z)$ for $R(M_Z)=1$ (left panel) and for 
randomly generated $R$ (right panel). 
\vspace*{0.cm}}
\label{f7}
\end{figure}

We have repeated the earlier analyses also for an inverted hierarchy of
light neutrino masses. Because the solar-neutrino mass scale is now higher
than the atmospheric one, the Yukawa couplings of the first generations
are generally larger. Therefore, all the effects related to the
renormalization of the solar parameters are generally enhanced. This can
be seen in Fig.~\ref{f7}, where we plot the neutrino mixing angles at
$M_{GUT}$ as functions of the lightest neutrino mass $m_{3}(M_Z)$ for
$R(M_Z)=1$ and for randomly-generated $R.$ If $R(M_Z)=1,$ the mixing
practically does not run even in the inverted-hierarchy case. The
exception is in the high-mass region $m_{3}(M_Z)\sim 0.1$ eV, when the
mass eigenvalues cross and the moderate degeneracy causes discrete jumps
in the mixing angles.  As expected, for the random choice of $R$, the
angle $\theta_{12}$ runs very strongly. The level-crossing stripe around
$\theta_{12}\sim 60^\circ$ exists for all values of $m_{3}$, while the
angles $\theta_{13}$ and $\theta_{23}$ run only for large values of
$m_{3}$.

\section{Discussion and Conclusions}

We have studied how the RGE running of neutrino mixing depends on the
unknown seesaw parameters. We have taken a bottom-up approach in which we
fix the known low-energy neutrino parameters to their measured values.  
Parametrizing the nine high-energy parameters of the seesaw mechanism via
the dominance matrix $R$, we have scanned over all the 18 seesaw
parameters by generating the unknown parameters randomly. The fact that
the matrix $R_{ij}$ measures the heavy neutrino $N_j$ contribution to the
light neutrino $\nu_i$ means that this parametrization is particularly
valuable for this study.  We have compared the results for random $R$
elements with the simple case $R = 1$.

We have found that significant running effects can occur only when some of
the light-neutrino masses have comparable contributions from more than one
heavy neutrino $N_i,$ and the light-neutrino mass scale is at least
moderately degenerate. For a normal hierarchy of neutrino masses, the
complementarity relation (1) between neutrino and quark mixing angles
evolves very little between the GUT scale and the electroweak scale.
However, the other oscillation angles $\theta_{13,23}$ run rather more
than $\theta_{12}$ and also beyond the expected measurement errors. In
certain cases, we observe level crossing in the light-neutrino mass
eigenstates that is accompanied by jumps in the oscillation angles. The
running of the CP-violating oscillation phase $\delta$ is
strong for random $R$ but insignificant for  $R = 1$.

The analysis presented here complements the top-down approach often
adopted elsewhere. It reveals some of the pitfalls in inferring properties
of the neutrino mass matrix generated at the GUT scale from current
low-energy measurements alone, in the absence of supplementary theoretical
or phenomenological input. We hope that these results may serve as useful
aids in the attempt to understand the neutrino mass matrix, which has
already revealed several surprises. The results presented here demonstrate
that our low-energy measurements are far from telling us the whole story.

\vskip1.4cm

\noindent {\bf Acknowledgment}\\

This work was supported by the ESF Grant 6140 and by the Ministry of
Education and Research of the Republic of Estonia. We thank S. Antusch for
useful communications.

\end{document}